\documentclass[runningheads]{svmult}

\usepackage{makeidx}   
\usepackage{graphicx}  
\usepackage{subeqnar}  
\usepackage{multicol}  
\usepackage{physprbb}  
\makeindex             

\begin{document}
\title*{The VVDS: early results on the large 
scale structure distribution of galaxies
out to $z\sim1.5$}
\toctitle{VVDS: early results on the large scale
\protect\newline  structures distribution to $z\sim1.5$}
%
%
\titlerunning{VVDS: early results on LSS distribution to $z\sim1.5$}
%
\author{O. Le F\`evre$^1$, G. Vettolani$^2$, D. Maccagni$^3$, 
J.P. Picat$^4$, 
C. Adami$^1$, M. Arnaboldi$^{5,12}$, S. Arnouts$^1$, S. Bardelli$^{10}$,
M. Bolzonella$^{10}$,
M. Bondi$^2$, D. Bottini$^3$, G. Busarello$^5$, A. Cappi$^2$, 
P. Ciliegi$^{10}$, T. Contini$^4$,
S. Charlot$^7$, S. Foucaud$^3$,  P. Franzetti$^3$, B. Garilli$^3$, 
I. Gavignaud$^4$, L. Guzzo$^8$, O. Ilbert$^1$, 
A. Iovino$^8$, V. Le Brun$^1$, B. Marano$^{11}$, C. Marinoni$^1$, 
H.J. McCracken$^2$, G. Mathez$^4$, A. Mazure$^1$,
Y. Mellier$^6$, B. Meneux$^1$
P. Merluzzi$^5$, , R. Merighi$^10$, S. Paltani$^1$,
R. Pell\`o$^4$, A. Pollo$^8$, L. Pozzetti$^2$,
M. Radovich$^5$, D. Rizzo$^8$, R. Scaramella$^9$, M. Scodeggio$^3$,  
L. Tresse$^1$, G. Zamorani$^{10}$, A. Zanichelli$^2$, E. Zucca$^{10}$ }
\authorrunning{Olivier Le F\`evre et al.}
%
%
\institute{$^1$Laboratoire d'Astrophysique de Marseille, France, 
$^2$Istituto di Radio-Astronomia - INAF, Bologna, Italy,
$^3$IASF - INAF, Milano, Italy, 
$^4$Laboratoire d'Astrophysique - Observatoire
Midi-Pyr\'en\'ees, Toulouse, France,
$^5$Osservatorio Astronomico di Capodimonte, Naples, Italy,
$^6$Institut d'Astrophysique de Paris, France,
$^7$Max Planck fur Astrophysik, Garching, Germany,
$^8$Osservatorio Astronomico di Brera, Milano, Italy,
$^9$Osservatorio Astronomico di Roma, Italy,
$^{10}$Osservatorio Astronomico di Bologna - INAF, Italy,
$^{11}$Dipartimento di Astronomia - Universita di Bologna, Italy,
$^{12}$Osservatorio Astronomico di Torino, Italy
}

\maketitle              

\begin{abstract}
The VIMOS VLT Deep Survey (VVDS) is an on-going program to 
map the evolution of galaxies, large scale structures and 
AGNs from the redshift measurement of more than 100000 objects
down to a magnitude $I_{AB}=24$, in combination with a 
multi-wavelength dataset from radio to X-rays. 
We present here the first results
obtained from more than 20000 spectra. 
Dedicated effort has been invested to successfuly enter
the ``redshift desert'' $1.5<z<3$, by producing high S/N
galaxy templates in the rest wavelength domain probed
by VIMOS.
We have produced the luminosity function (LF) of galaxies
for the first time out to $I_{AB}=24$, and the evolution
of the LF for each morphological type is clearly established,
indicating that most of the LF evolution is concentrated
in the later types.
We present for the first time the 3D spatial
distribution of a well controlled sample of faint
galaxies, and show that significant
clustering is detected out to $z\sim1.5$.  
\end{abstract}

\section{Introduction}

The VIMOS VLT Deep Survey is dedicated to study galaxies, large scale 
structures and AGN evolution up to $z\sim5$. 
To determine the evolutionary status
of the galaxy and AGN population as a function of lookback time, it
has been recognized early in the survey design phase that
the large range in redshifts and the complex physical phenomena probed 
require to expand the core observations of the VVDS from the
visible domain to a broad
multi-wavelength spectrum. Observing the rest-frame SED of 
galaxies (e.g. in the B-band rest frame) at the same wavelength 
from $z\sim0$ to $z\sim5$, and having a sufficiently accurate
estimate of the k-correction necessary to compute absolute magnitudes, 
requires to obtain deep photometric data from the U band to the K band. 
Furthermore, to get a complete picture of e.g. the star formation
process, for galaxies with 
variable amount of dust extinction, or to study the evolution
of large scale structures or AGNs, 
it is necessary to expand the wavelength baseline
of observations from the far UV to the far IR, from the radio-waves to
the X-rays. 


We describe here the multi-wavelength approach followed by the
VVDS, and present early results from the first epoch 
VVDS observations. We describe the effort to assemble for
the first time an unbiased magnitude limited 
sample of galaxy redshifts and spectra $I_{AB}\leq24$
up to $z\sim5$, going through the
``redshift desert'' $1.5<z<2.5$,  and present 
preliminary results on the evolution of galaxies and large scale structures
up to $z\sim1.5$.

\section{The multi-wavelength VVDS}
The VIMOS VLT Deep Survey has been designed from the start
as an extensive survey with multi-color BVRI data obtained 
at CFHT \cite{olf1}, and deep 
spectroscopy obtained with VIMOS
on the VLT \cite{olf3}, forming the core of the survey.  
The core observations are obtained on
several fields, each $2\times2$deg$^2$: 
(i) multi-color photometry from B to I band, leading to clean I-band magnitude 
selected samples and broad band spectral energy distributions (SED), 
and (ii) multi-object spectra with a goal of $\sim100000$ objects
to $I_{AB}=22.5$ and $I_{AB}=24$. 

In addition, our VVDS team has been 
conducting observations in other wavelength
(U and K bands, radio), and has been connecting 
with other survey groups to 
stimulate other wavelengths observations in the VVDS
fields (Galex-UV, 
X-rays with XMM and Chandra, SIRTF, see these proceedings).

We have also been advocating the need for the accurate
knowledge of galaxy morphologies over large scales with the HST,
and we are participating to the HST-COSMOS program
set to observe 2deg$^2$ on the VVDS-10h field with the
Advanced Camera for Surveys (ACS) \cite{scoville}.

The current spectroscopic VVDS data has been obtained in 
November- December 2002
in an observing campaign from which 18  nights were clear.
More than 20000 spectra have been obtained, from which about 
15000 secure galaxy and AGN redshifts are available in 3 different fields.

\section{Accessing the ``redshift desert'' $1.5<z<2.5$}
The VVDS is the first survey to assemble  a 
complete sample of galaxies based on
a simple I-band limit down to $I_{AB}=24$. 
Magnitude limited samples have
the advantage of a controlled bias in the selection
of the target galaxies, which can lead to a 
secure census of the galaxy population  
as seen at a given rest-frame wavelength (see e.g.
\cite{lilly}). The drawback 
is that, as the magnitudes get fainter, the redshift
range gets larger, and identifying redshifts
out of a very large range of possibilities becomes
increasingly harder from a fixed set of observed
wavelengths.

The VVDS default wavelength range for the VIMOS observations
is $5500-9500\AA$, observed at a spectral resolution 
$R\sim200$ to allow for a large multiplex gain 
during the multi-slit VIMOS observations. 
It allows a secure 
follow-up of the spectral signature of galaxies 
from [OII]3727$\AA$ to the 4000$\AA$
region and minimize the bias in the identification of galaxy 
redshifts up to $\sim1.5$. However,
measuring the redshifts of galaxies with
$1.5<z<3$ is quite challenging with generally faint
features, and a lack of published observed
galaxy templates in the range $1800-3000\AA$
to be used in cross-correlation programs
such as the KBRED environments developed for the
VVDS \cite{scara}. Breaking up
this ``redshift desert'' is critical to reduce the
incompleteness of deep redshift surveys, and probe
the galaxy population at an important time in the
evolution.

In a first pass of redshift measurements of the
first epoch spectra of the 
VVDS $I_{AB}=24$ sample \cite{olf2}, 
the redshift of $\sim$10\% of the galaxies appeared hard
to identify while a significant fraction of these spectra
had a good continuum S/N. We have setup an iterative
approach to identify these objects, based on the 
creation of rest-frame galaxy templates going 
as far as possible toward the UV, bridging the gap
between the well observed wavelength
range above [OII]3727$\AA$ and the range below 
1800$\AA$ with the templates assembled by Shapley
et al. \cite{shapley}. These new templates 
(see Figure 1) have been included in 
our redshift measurement engine KBRED
and tested against the parent samples of galaxies used to produce
them, in the range up to redshift $z=1.5$,
then up to redshifts $z=2$ and above as new redshifts were being
identified.  

This approach has been quite successful
on test samples, successfully turning up  galaxies
at redshifts $1.5<z<2.5$ as shown in Figure 2. We are currently
going over our data with the expectation to significantly
reduce the incompleteness which was observed at the end of our 
first pass.

\begin{figure}[h]
\begin{center}
\includegraphics[width=0.5\textwidth,angle=-90]{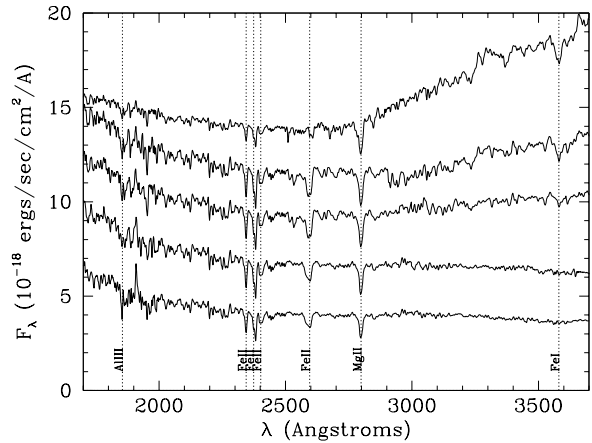}
\end{center}
\caption[]{Blowup of the VVDS galaxy templates in the range $1700-3700$\AA,
produced from the combination of high redshift galaxies observed in the range
$1.3-4$ with VIMOS}
\label{eps1}
\end{figure}

\begin{figure}[h]
\begin{center}
\includegraphics[width=0.5\textwidth,angle=-90]{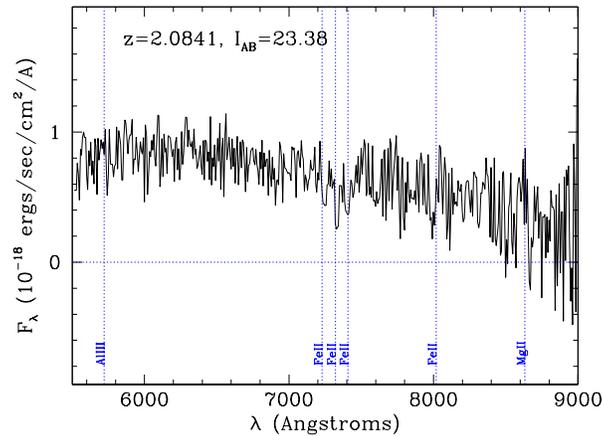}
\end{center}
\caption[]{VVDS galaxy identified in the ``redshift
desert'' at a redshift z=2.0841}
\end{figure}

\section{Galaxy evolution}
A preliminary description of the evolution of the 
galaxy luminosity function out to $z=1.5$
has been presented in Le F\`evre et al. \cite{olf2}.
The current VVDS data allows to build e.g. the
luminosity function and trace its evolution accross
many time steps and, furthermore, to break up the
population vs. spectral type, or vs. the local galaxy
density. 

In Figure 3, we present the evolution of the LF of galaxies
out to redshift 1.5, for each of 4 spectral types from
early to late types. Most of the LF evolution is concentrated
in the later types, very strong for the type 4 presented
in Figure3, while the LF of early types 
does not seem to change much over this period. 
These results, together with the Luminosity Density
evolution will be discussed 
extensively in forthcoming papers \cite{ilbert},\cite{zucca},
\cite{tresse}.
The evolution of the star formation rate from U to K band data
will be compared with the star formation indicators obtained
at other wavelengths \cite{tresse}, e.g. in the radio \cite{bondi},
or in the far UV with Galex \cite{martin}.

\begin{figure}[h]
\begin{center}
\includegraphics[width=0.9\textwidth,angle=0]{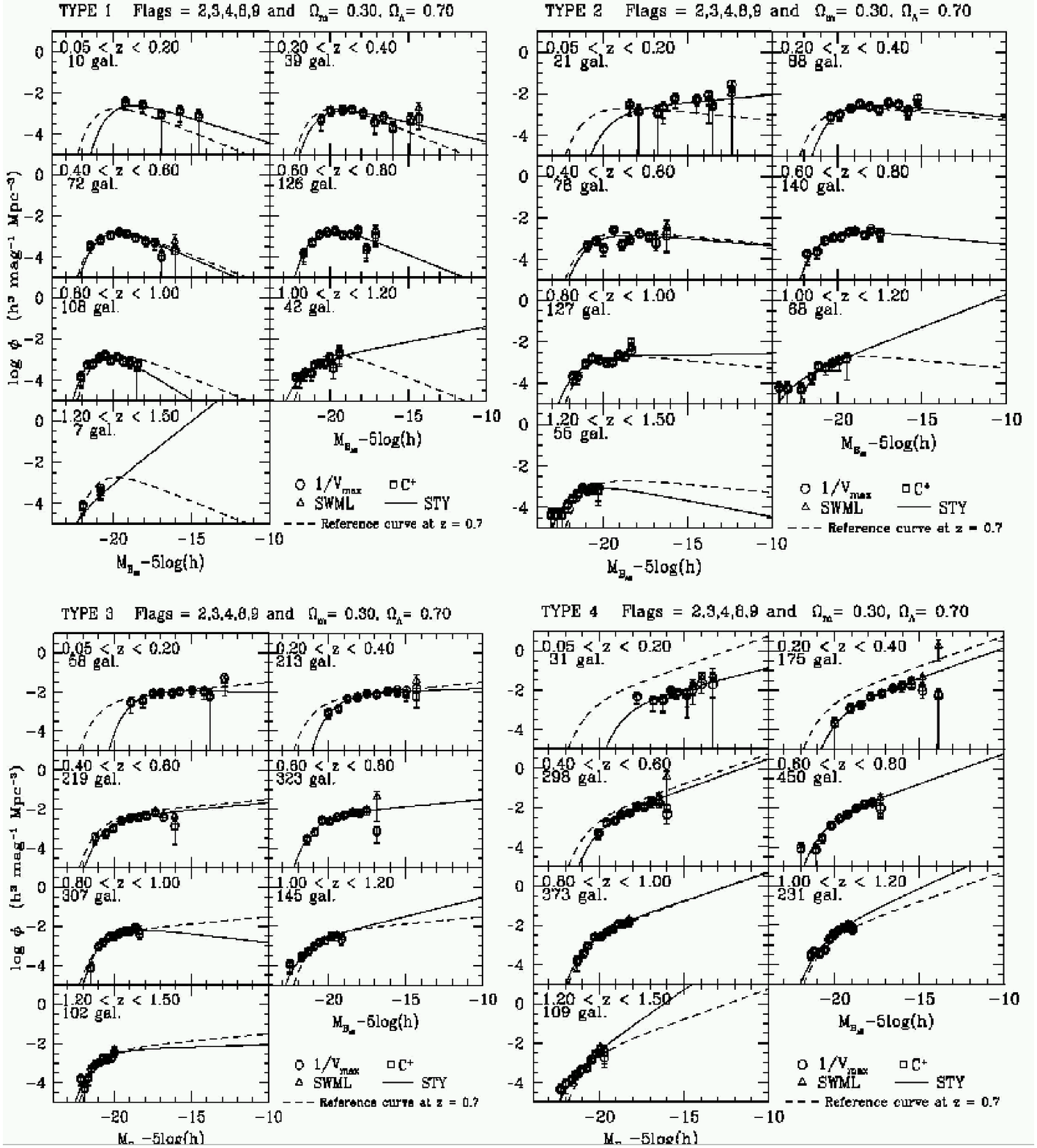}
\end{center}
\caption[]{Evolution of the galaxy luminosity function from the 
$I_{AB}\leq24$ VVDS sample, separated in 4 spectral types from early types 
(upper left) to late types (bottom right). Most of the
LF evolution is concentrated in the later type (preliminary,
courtesy O. Ilbert).}
\end{figure}

\section{Large scale structure evolution}
The VVDS has been designed to trace large
scale structures out to $\sim100$Mpc at $z\sim1$.
Although our current data from the 
on-going survey does not yet reach these 
scales, we have assembled a set of redshifts
in an area $\sim30\times30$arcmin$^2$,
with $\sim1/3$ of all galaxies to $I_{AB}=24$
with a measured redshift, which allows
a first pass at tracing the large scale structures.

The preliminary redshift histogram of the VVDS-0226-04
field, shows a dense 
alternance of strong peaks in the 1D projection (Figure 4). 
To further detail the 3D distribution of galaxies,
we have derived the galaxy density field
from this data set using the Wiener
filtering technique with a 3 Mpc smoothing scale. The deep probe 
in this field extends over more than 3000 Mpc,
for a total volume sampled of $\sim3\times10^6$Mpc.
Several slices from this deep probe are shown in Figure 5.
The density field appears to be 
already significantly clumped by $z\sim1$ 
\cite{marinoni}. As an example, we show in Figure 6
a strong overdensity and very complex structure 
at $z\sim0.962$ spreading accross
the full $30\times30$Mpc$^2$ of the field. 

\begin{figure}[h]
\begin{center}
\includegraphics[width=0.6\textwidth]{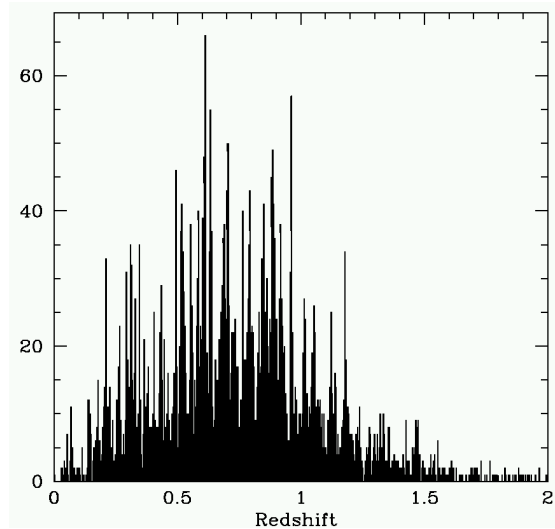}
\end{center}
\caption[]{Redshift distribution of galaxies in the
field VVDS-0226-04. A total of 5010 galaxies are represented
here from the $I_{AB}\leq24$ magnitude limited sample. Strong 
galaxy peaks are traced all the way to $z\sim1.5$. }
\end{figure}

\begin{figure}[h]
\begin{center}
\includegraphics[width=1.\textwidth]{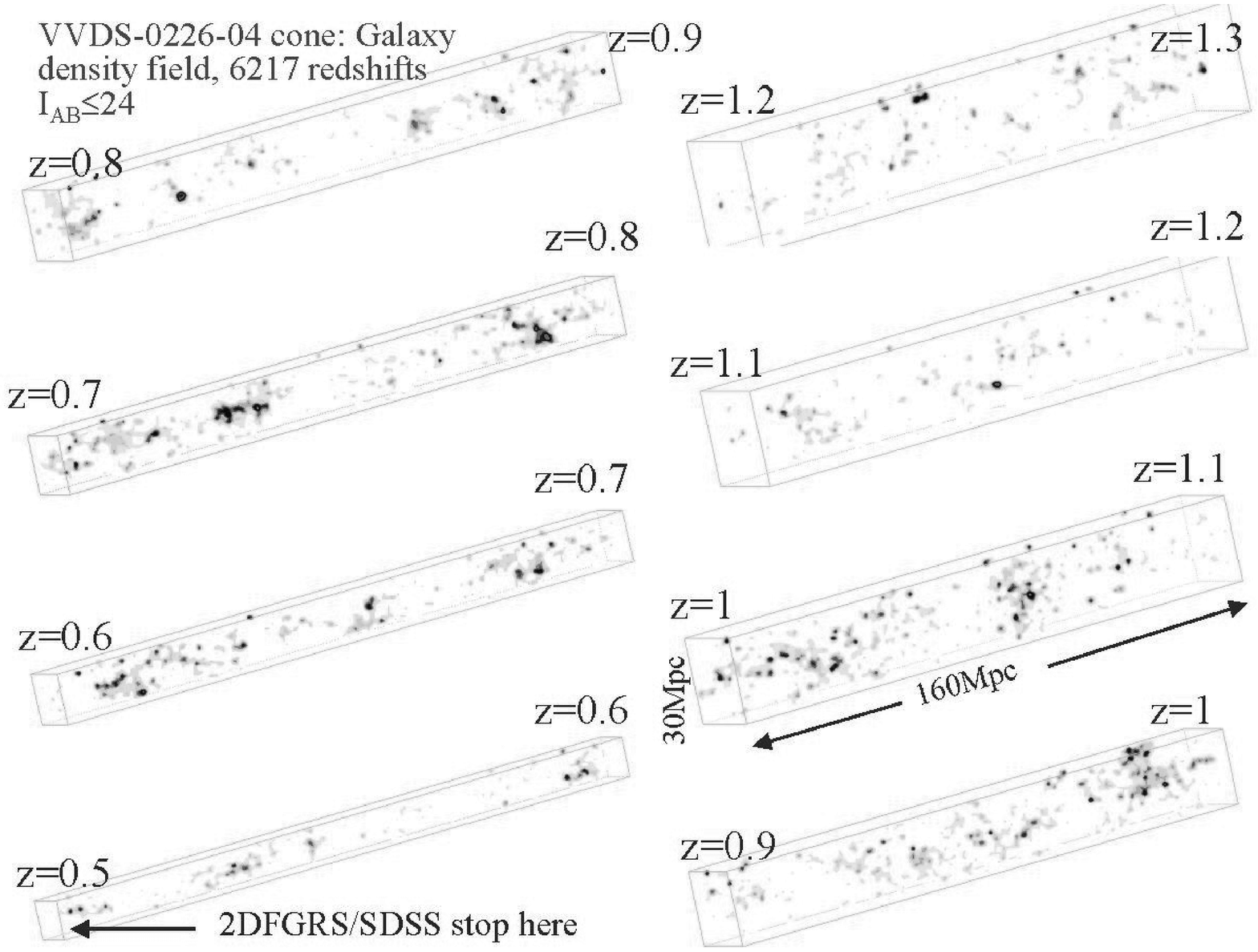}
\end{center}
\caption[]{Galaxy density field in the VVDS 0226-04 field in 
several redshift boxes up to $z=1.3$ (preliminary, courtesy
C. Marinoni). }
\end{figure}

\begin{figure}[h]
\begin{center}
\includegraphics[width=0.8\textwidth]{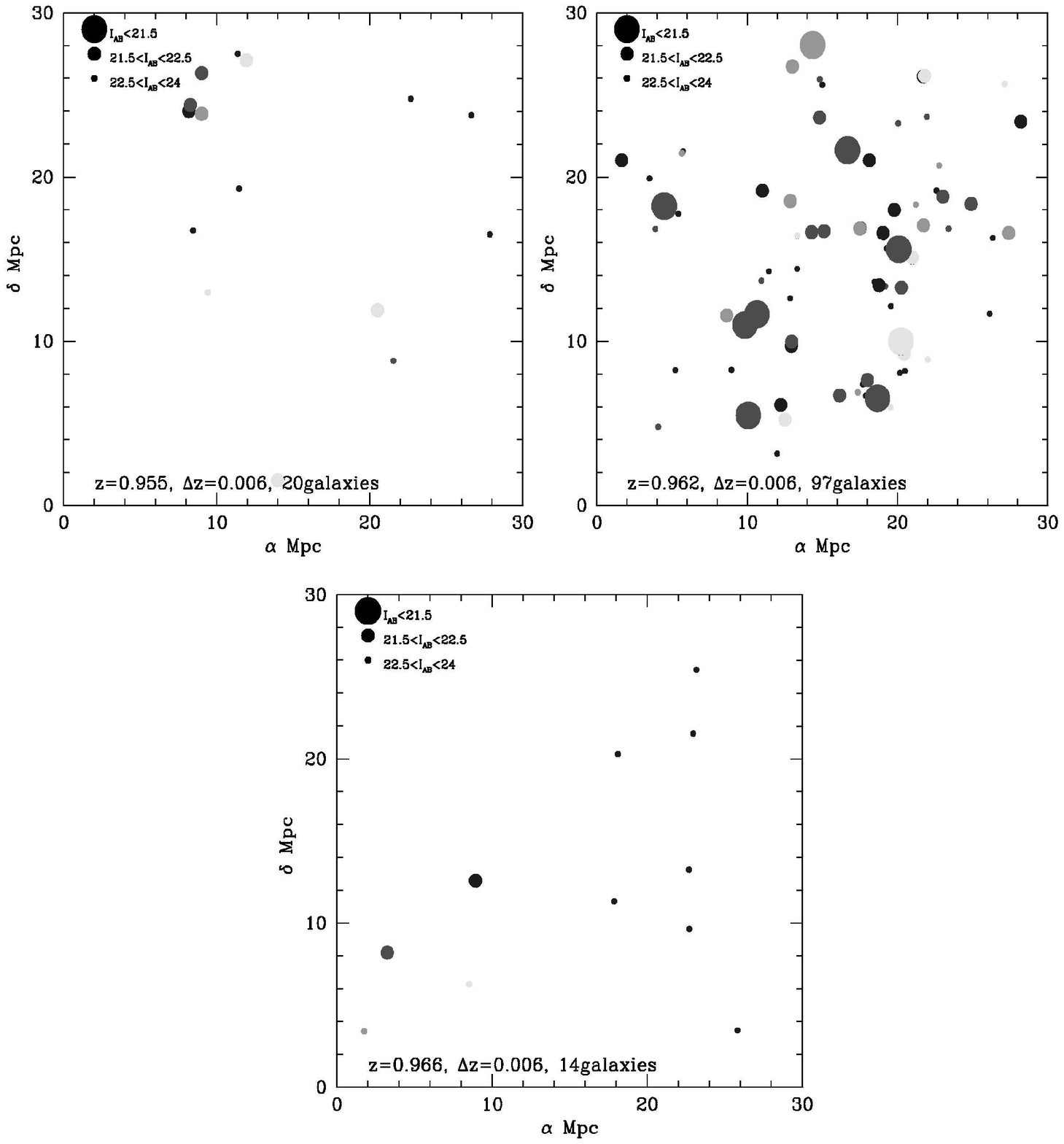}
\end{center}
\caption[]{Strong galaxy overdensity observed in the 
VVDS 0226-04 field, from left to
right: galaxies observed in a $\Delta z=0.006$ slice at $z=0.955$, 
$z=0.962$, and $z=0.966$. The central slice has nearly 100 galaxies
observed covering the full $30\times30$Mpc$^2$ in a web-like
distribution, while the adjacent slices have only 14 and 20 galaxies
respectively.}
\end{figure}

The VVDS 0226-04 field has been observed by XMM \cite{pierre}, 
this will provide the opportunity to cross-correlate the 
X-ray flux from hot gas in
clusters and groups with the structures identified from the
redshift measurements.

The VVDS field at 1000+03 is now the focus of an extensive 
observational effort as part of the HST-COSMOS program
\cite{scoville}, including the measurements of more than
50000 redshifts in the $2deg^2$ area to trace the galaxy density field.
The other VVDS survey fields at 1400+00 and  2217+00 will continue
to be the target of the VVDS team in future observational
campaigns to trace the large scale structure distribution out
to $z\sim1.5$ in several independent fields.  

\section{Summary}
The VVDS is providing for the first time a magnitude
limited sample of $\sim15000$ galaxy redshifts, with $\sim11000$
redshift in the $I_{AB}\leq24$ sample, and $\sim4000$ in the
$I_{AB}\leq22.5$ sample,
going successfuly through the ``redshift desert'' $1.5<z<2.5$.
The evolution of the luminosity function up to
$z\sim1.5$ is traced for several spectral type,
with most of the evolution concentrated in the later types.
Strong overdensities in  the distribution of galaxies
are visible over scales in excess of 30Mpc up to
$z\sim1.5$. This provides evidence that the galaxy distribution was already
strongly structured when the universe was half of its
present age.\\

Acknowledgements: We thank C. Cesarsky and ESO for the early allocation
of VLT Garanteed Observing Time, and A. Shapley who has provided fits
files for her nice UV templates for $\lambda\leq1700\AA$.

\end{document}